\documentclass[aps, prb, reprint, nofootinbib, longbibliography,floatfix, superscriptaddress]{revtex4-2}

\usepackage[utf8]{inputenc}
\usepackage{float}
\usepackage{graphicx}  
\usepackage{physics}
\usepackage{esint}
\usepackage{amssymb}  
\usepackage{mathtools}
\usepackage{amsmath}
\usepackage{amsthm}
\usepackage{mathrsfs}
\usepackage[dvipsnames]{xcolor}

\usepackage[colorlinks=true,linkcolor=purple,citecolor=purple, urlcolor=purple,plainpages=false,pdfpagelabels]{hyperref}
\usepackage{color,xcolor,colortbl}
\usepackage{bm}

\newcommand{\dblval}[1]{\left\langle\!\left\langle#1\right\rangle\!\right\rangle}

\numberwithin{footnote}{section}
\numberwithin{equation}{section}

\begin{document}

\title{A numerical study of measurement-induced phase transitions in the Sachdev-Ye-Kitaev model}

\author{Stav Haldar}\email{hstav1@lsu.edu}\affiliation{Department of Physics \& Astronomy, Louisiana State University, Baton Rouge, LA 70803, USA}
\author{Anthony J. Brady}\email{ajbrady4123@arizona.edu}
\affiliation{
Department of Electrical and Computer Engineering, University of Arizona, Tucson, Arizona 85721, USA
}

\date{\today}



\begin{abstract}

Continuous monitoring of an otherwise closed quantum system has been found to lead to a measurement-induced phase transition (MIPT) characterized by abrupt changes in the entanglement or purity of the many-body quantum state. For an entanglement MIPT, entangling dynamics compete with measurement dynamics, pushing the system either to a phase with extensive entanglement or to a phase with low-level entanglement. For purification MIPTs, projective measurements effectively cool and localize the system, inducing a transition from a mixed state to an uncorrelated pure state. In this work, we numerically simulate monitored dynamics in the all-to-all Sachdev-Ye-Kitaev (SYK) model for finite $N$. We witness both entanglement and purification MIPTs in the steady-state. It is often said that there is an equivalence between entanglement and purification MIPTs, however we provide numerical evidence to the contrary, implying that entanglement and purification MIPTs are indeed two distinct phenomena. The reason for such a distinction is quite simple: entanglement can revive after a completely projective measurement---if measurements do not occur too often in time---but impurity cannot.
\end{abstract}

\maketitle




\section{Introduction}

Quantum statistical mechanics accurately captures most properties of many-body quantum systems at equilibrium where, e.g., the system can be described by a handful of macroscopic quantities, such as the temperature, total particle number etc. However, many aspects of quantum many-body systems out of equilibrium~\cite{eisert2015quantum}---such as entanglement growth in closed quantum systems~\cite{calabrese2005evolution,nahum2017ent_growth,sondhi2018hydro}---require more care. Tools from quantum information have been employed to better understand, characterize, and even construct highly correlated quantum many-body states (or quantum matter) in and out of equilibrium~\cite{vedral2008manybody,zeng2019qit_qm, lu2022meastoQM}. 

The dynamics of a closed quantum system out of equilibrium is governed by a Hamiltonian that introduces correlations (entanglement) between the system's constituents. Starting from an initially uncorrelated state, the entanglement in the system evolves in time, eventually saturating to some finite value; in which case, the steady-state is a many-body entangled state. If the interactions are strong and the subsystems are highly connected, then the growth of entanglement is fast compared to all other time scales, and entanglement saturates to an extensive value. Circumstances change when we ``open'' the system and externally perturb it with incoherent probes (i.e., non-unitary perturbations), such as coupling the system to a heat bath or measuring parts of the system. 

\begin{figure}
    \centering
    \includegraphics[width=.9\linewidth]{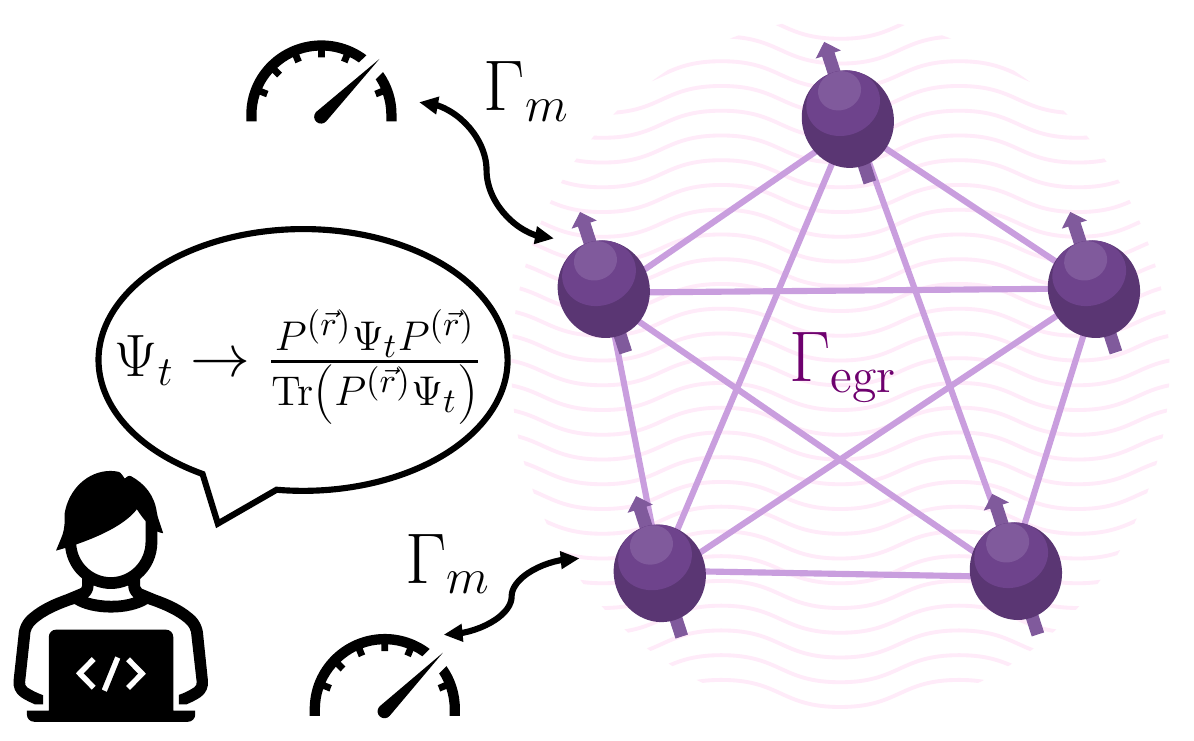}
    \caption{Projective measurements sporadically interrupt the internal, unitary time evolution of a quantum system (here, 5 spins with all-to-all connectivity). A competition between the measurement rate $\Gamma_m$ and the entanglement growth rate $\Gamma_{\rm egr}$ determines the entanglement dynamics of the system.}
    \label{fig:syk_measure}
\end{figure}

In this work, we focus on continuously probing a quantum system with measurement devices that we have access to---a process called continuous monitoring. In particular, we assume that the internal dynamics of the system are spontaneously interrupted by local projective measurements characterized by a measurement strength $p_m$ and a measurement rate $\Gamma_m$; see Fig.~\ref{fig:syk_measure} for an illustration. Such monitoring dynamics have been feverishly studied in the past several years, primarily via brickwork circuit models~\cite{graeme2019projective,li2018zeno,skinner2019MIPT,fisher2019driven,szy2019weakMeas,jian2020criticality,gullans2020criticality,bao2020theoryof} (see also the recent reviews~\cite{potter2022rvw,fisher2022rvw} and references therein) which intersperse discrete two-body interactions with local projective measurements. 

In brickwork circuit models~\cite{graeme2019projective,li2018zeno,skinner2019MIPT,fisher2019driven,szy2019weakMeas,jian2020criticality,gullans2020criticality,bao2020theoryof,potter2022rvw,fisher2022rvw}, it has been shown that a competition arises between the entanglement dynamics of the closed system, which drives the system into a many-body entangled state, and the decoupling dynamics induced by continuous monitoring, which drives the system in an uncorrelated product state. This competition leads to different phases of matter and to a so-called measurement induced phase transition (MIPT), depending on whether the internal entangling dynamics or the measurement dynamics dominates the evolution. Similar studies with (random) Hamiltonian evolution~\cite{cao2019fermion,swingle2021prl,swingle2021miptBrownian,swingle2022longrangeMIPT,liu2021syk_chain,Turkeshi2021ising,pichler2022entanglemenTraj}---which include random Brownian circuits (continuous-time analogs of brickwork circuit models)~\cite{swingle2021miptBrownian,swingle2022longrangeMIPT,pichler2022entanglemenTraj} and large $N$ analytically solvable models of coupled clusters of SYK chains ~\cite{swingle2021prl,liu2021syk_chain}---have come to similar conclusions. A numerical study of the effects of decoherence on information scrambling and growth of entanglement for several many body quantum Hamiltonians including the SYK model was also performed in ~\cite{Touil2021}. Recent experiments regarding entanglement growth and MIPTs in brickwork circuits with Noisy Intermediate-Scale Quantum (NISQ~\cite{preskill2018nisq}) devices have also been performed~\cite{google2021scrambling,noel2022observation,koh2022experiment}.

MIPTs come in two flavors---an entanglement MIPT and a purification MIPT---which are often conflated into the singular phenomenon of a MIPT, however we later discuss how these two phenomena are in fact distinct. An entanglement MIPT refers to the transition from a quantum state with extensive entanglement (volume-law phase) to a quantum state with sub-extensive entanglement (area-law phase). Whereas a purification MIPT refers to the transition from a highly mixed quantum state (mixed phase) to an uncorrelated pure state (pure phase).\footnote{Another intriguing perspective on purification MIPTs comes from the theory of quantum error correction~\cite{gullans2020purification,choi2020qec_prl,gullans2020scalable,ippoliti2021postselection,fan2021self,gullans2021low_depth,li2021SM_qec,hastings2021qumems}, whereby interprets the open quantum system as a quantum memory that is robust to decoherence due to measurements.} For discrete brickwork circuit models~\cite{li2018zeno,skinner2019MIPT,fisher2019driven,szy2019weakMeas,jian2020criticality,gullans2020criticality,bao2020theoryof,potter2022rvw,fisher2022rvw}, the critical point marking a MIPT depends on the probability that a local measurement will occur as well as the circuit depth prior to a measurement round~\cite{choi2020qec_prl}. For physical systems evolving under a Hamiltonian~\cite{pichler2022entanglemenTraj}, the measurement strength and the rate of entanglement growth in the closed system establishes a dynamical rate that competes against the local measurement rate and dictates which phase the system is in (as well as the corresponding critical points). 

In this work, we numerically study entanglement and purification MIPTs in the all-to-all SYK model~\cite{sy1993syk,kitaev2015syk,sachdev2022rmp} with finite $N$, where $N$ is the number of Majorana fermions (here, $N$ ranges from 10 to 20). We simulate monitored dynamics and track the entanglement or purity of the resulting quantum trajectories through time. We witness entanglement and purification MIPTs in the steady-state and observe a clear distinction between the two phenomena.

\section{Entanglement growth in the SYK model}

The SYK model~\cite{sy1993syk,kitaev2015syk,sachdev2022rmp} is a strongly interacting model for many-body quantum systems without any quasiparticle excitations. Low energy equilibrium states and dynamics of the system cannot be described in terms of quasiparticle excitations, as is the case for standard Fermi liquids, and even the ground state is an entangled quantum many-body state. Such systems are important from several different perspectives. From the point of view of condensed matter physics, such models provide a window into the fascinating world of strange metals and high temperature superconductors which are yet to be fully understood~\cite{sachdev2022rmp}. 

A feature that underlies many interesting aspects of the SYK model (and other strongly interacting models without quasiparticles) is the fast scrambling of information. From a physical point of view, fast scrambling can be understood as a faster-than-usual equilibration of the system\footnote{As compared to a weakly interacting system with well defined slow moving quasiparticles that do not collide often. Note equilibration times in Fermi liquids diverges for low temperatures.} after a local perturbation. This perspective is also important for understanding entanglement dynamics and effect of measurements on it. Fast scramblers such as the SYK system quickly regenerate their extensive, steady-state entanglement after a short lived local perturbation (like a projective measurement) occurs on a few sites. 

Consider $N$ all-to-all interacting Majorana fermions, where the each interaction term includes 4 sites. The coupling constants are site dependent random variables with zero mean $\langle\mathcal{J}_{ijkl} \rangle = 0$ and finite variance $\langle \mathcal{J}_{ijkl}^2 \rangle = 6J^2/N^3$, where $J$ defines the strength of the interactions. Let $\chi_i$ be the second-quantized Majorana field operator at the site $i$. The SYK Hamiltonian is then
\begin{eqnarray}
    H_{\mathcal{J}} = \sum_{1 \leq i<j<k<l\leq N} -\mathcal{J}_{ijkl} \, \chi_i \chi_j \chi_k \chi_l.
    \label{eqn:Hamiltonian}
\end{eqnarray}
In order to simulate the evolution of states under this Hamiltonian for finite $N$, we use exact diagonalization with the associated limitation of not being able to simulate $N > 24$. Although techniques such as random matrix models or approximate diagonalization methods like the density matrix renormalization group allow the handling of larger number of sites, they do not provide the full spectrum of the Hamiltonian, which is important for the dynamics of strongly interacting systems. 

We change our basis from Majorana operators $\chi$ to spin-$1/2$ Pauli operators $\{\sigma^x, \sigma^y, \sigma^z\}$ for computational purposes. This is done by the standard Jordan-Wigner transformation. Since two Majoranas map to one fermion, odd and even Majoranas are related to Pauli string operators as 
\begin{eqnarray}
    \chi_{2i-1} &=& \frac{1}{\sqrt{2}} \sigma^x_1 \sigma^x_2 ... \sigma^x_{i-1} \sigma^z_i, \\
    \chi_{2i} &=& \frac{1}{\sqrt{2}} \sigma^x_1 \sigma^x_2 ... \sigma^x_{i-1} \sigma^y_i.
\end{eqnarray}

As was briefly mentioned above, the fast scrambling behaviour of an all-to-all model like the SYK can be described via entanglement dynamics. More specifically let us start by looking at the rate at which entanglement grows in the system under the action of the SYK Hamiltonian and without any measurements. We measure the entanglement using half-chain entanglement entropy. Consider the system of $N/2$ spin-$1/2$ particles ($N$ Majorana fermions) on a linear chain with a fermion on each site. Also, consider partitioning the chain into two halves $A$ and $B$, such that $\log\abs{A}=\log\abs{B}=N/4$. We start with a unentangled product state at $t=0$. The initial state is an all-up state $\Psi_{AB}(0)= \ket{\Psi(0)} \bra{\Psi(0)}$ where $\ket{\Psi(0)} = \ket{1}_1\ket{1}_2...\ket{1}_{N/2}$. Here, $\ket{1}$ is the eigenstate of $\sigma^z$ with eigenvalue $+1$ ($\ket{0}$ is the eigenstate with eigenvalue $-1$). The SYK Hamiltonian is then ``switched-on'' such that the state at time $t$ is
\begin{equation}\label{eq:rho_HJt}
    \Psi_{AB}(t;\mathcal{J}) = e^{-iH_{\mathcal{J}}t/\hbar} \Psi_{AB}(0) e^{+iH_{\mathcal{J}}t/\hbar}.
\end{equation}
We have included the label $\mathcal{J}$ into the argument of the state since $\Psi_{AB}(t;\mathcal{J})$ corresponds to a single realization $H_{\mathcal{J}}$. 

\begin{figure}
    \centering
    \includegraphics[width=.9\linewidth]{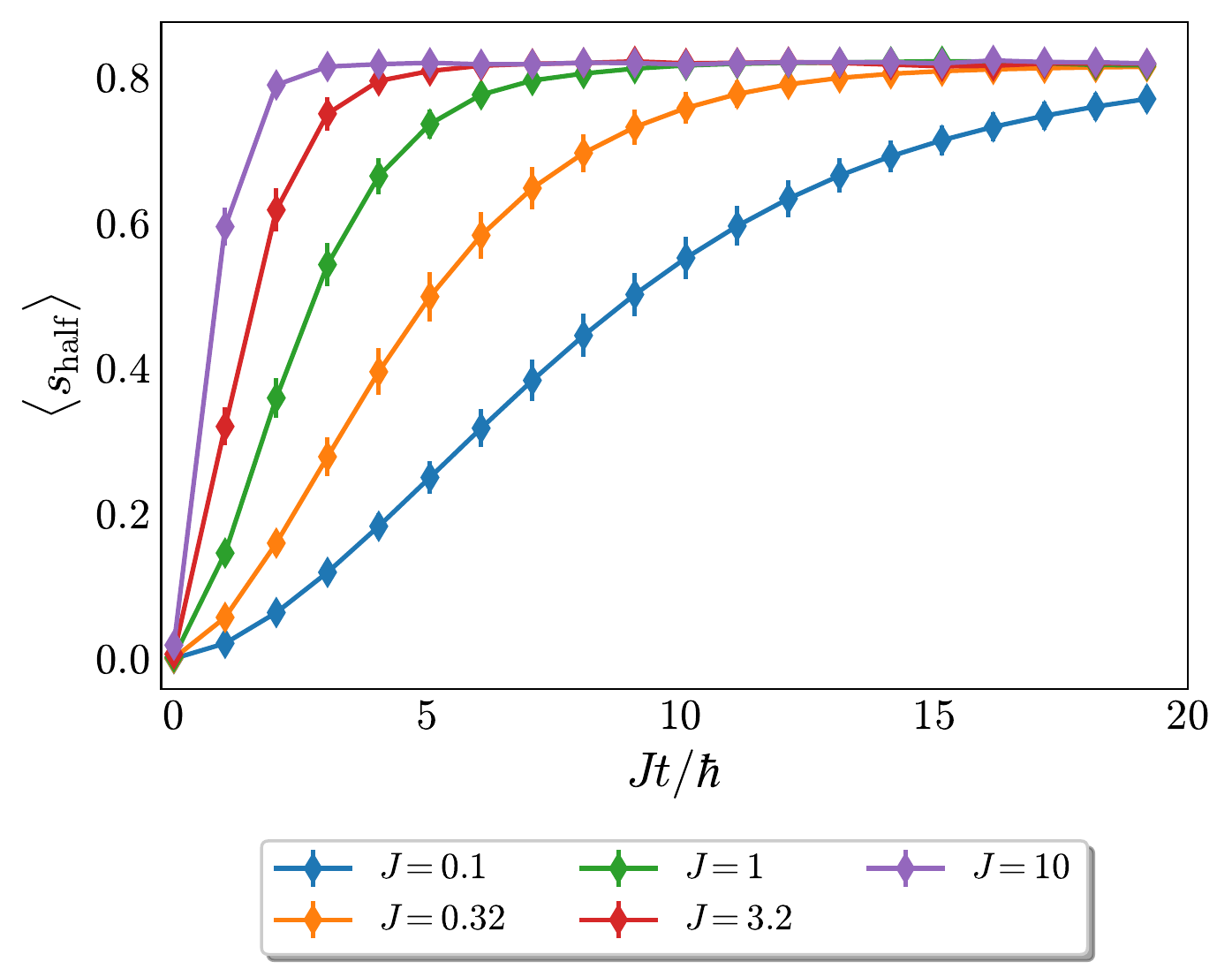}
    \caption{Entanglement entropy (ensemble averaged) in a SYK chain with $N=16$ Majoranas, as a function of time. The coupling strength $J=1$. The dynamics have been averaged over 50 runs. }
    \label{fig:S_t}
\end{figure} 

The half-chain entanglement entropy of the state $\Psi_{AB}$ at time $t$ is,
\begin{equation}
    S_{\rm{half}}(t;\mathcal{J}) =- \Tr(\Psi_A\log{\Psi_A})
\end{equation}
where $\Psi_A\equiv \Tr_B(\Psi_{AB}(t;\mathcal{J}))$. We normalize the entropy by the number of particles in the half-chain ($N/4$) and define the half-chain entropy density $s_{\rm half}\equiv 4S_{\rm half}/N$, such that $0\leq s_{\rm half}\leq1$. Since $\mathcal{J}$ is a random variable, we average over many Hamiltonian realizations $H_{\mathcal{J}}$ to compute the ensemble averaged entropy at time $t$,
\begin{equation}
   \expval{s_{\rm half}(t)}_{J}\equiv\int\dd{\mathcal{J}}p(\mathcal{J})s_{\rm{half}}(t;\mathcal{J}),
\end{equation}
where $p(\mathcal{J})$ is a zero-mean Gaussian distribution for the random coupling strength $\mathcal{J}$. Due to the ensemble average and the nature of the distribution $p(\mathcal{J})$, the average half-chain entropy depends only on the standard deviation $J$.  Figure \ref{fig:S_t} shows the evolution of the half-chain entanglement entropy as a function of time for different values of interaction strength $J$ for $N=16$ Majorana fermions. For stronger interactions (higher $J$), entanglement growth is faster, and the saturation value of $\expval{s_{\rm half}(t\rightarrow\infty)}_{J}\approx.8$ (less than 1 due to finite $N$) is approached more quickly.

\begin{figure}
    \centering
    \includegraphics[width=.9\linewidth]{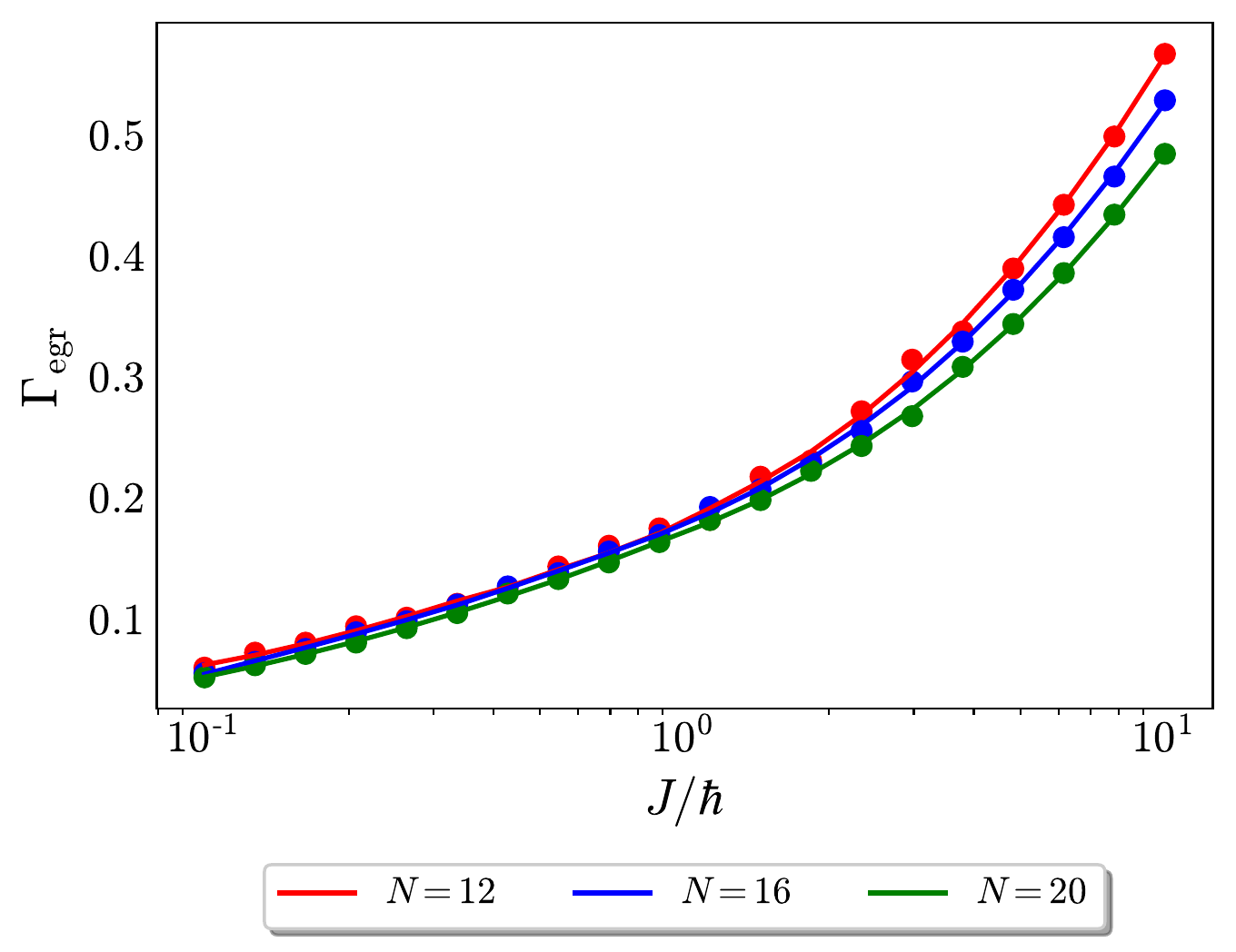}
    \caption{Average entanglement growth rate $\Gamma_{\rm egr}$ as a function of $J$, for different values of $N$. Dynamics have been averaged over 50 runs.}
    \label{fig:EGR}
\end{figure}

MITPs arise due to a competition between internal scrambling of the system (generated by the many-body Hamiltonian $H_{\mathcal{J}}$) and the rate of measurements. It is thus important to quantify how quickly many-body correlations build up within the system, which we quantify by a so-called entanglement growth rate (EGR). We explicitly define the EGR as the rate of change of entanglement starting from a product state in the absence of measurements, 
\begin{eqnarray}
    \rm{EGR}(t,J) \equiv \dv{\expval{s_{\rm half}(t)}_{J}}{t}.
\end{eqnarray}
The EGR is a function of time, as is clear from Figure \ref{fig:S_t}. For instance, EGR is larger for small times (near $t=0$) and slowly falls to $0$ as steady state is reached. Thus in order to quantify the EGR in a time-independent way, we define a time average EGR,
\begin{eqnarray}
    \Gamma_{\rm egr} \equiv\frac{1}{\Delta t}\int_{\Delta t}\dd{t}\rm{EGR}(t,J).
    \label{eqn:gamma_egr}
\end{eqnarray}
The time average is taken over a period $\Delta t=t_{3/4}-t_{1/4}$, where $t_{3/4}$ and $t_{1/4}$ are implicitly defined by the following relations: $s_{\rm half}(t_{1/4})\equiv \expval{s_{\rm{half}}(\infty)}_J/4$ and $s_{\rm{half}}(t_{3/4}) \equiv 3\expval{s_{\rm{half}}(\infty)}_J/4$. Here, $\expval{s_{\rm{half}}(\infty)}_J$ is the saturation value of the half-chain entanglement entropy; e.g., $\expval{s_{\rm{half}}(\infty)}_J\approx.8$ for $N=16$ and for all values of $J$ (see Fig.~\ref{fig:S_t}).

In Fig.~\ref{fig:EGR}, we plot $\Gamma_{\rm egr}$ versus the coupling strength $J$ for different values of $N$. The EGR grows monotonically (and non-linearly) with the coupling strength $J$ but is nearly independent of $N$. The $N$ independence is owed to the normalization in the Hamiltonian [see discussion surrounding Eqn. \eqref{eqn:Hamiltonian}]. Note that the coupling strength $J$ sets an effective time-scale for the interactions whereas $\Gamma_{\rm egr}$ sets an effective time-scale for the growth of many-body correlations which develop at a slower rate.

\section{Measurement dynamics}

We give a description of the monitored dynamics of the SYK system. The measurements are done in the $\sigma_z$ basis at each site. Further, the measurements are independent Markov processes in the sense that the probability that a measurement takes place at a particular site and at a particular time is independent of the measurements on any other site or on the measurement history of the site itself. 

For every time step we make a binary choice of whether to make a measurement or not. This decision is taken through a Monte Carlo method. In other words, generate a random number $r$ between 0 and 1; if $r \leq r_m$, perform a measurement, else do not. Here $r_m$ is determined by the measurement rate $\Gamma_m$,
\begin{eqnarray}
    r_m = \Gamma_m dt,
    \label{eqn:gamma_m}
\end{eqnarray}
where $dt$ is the simulation time step. The above equation can also be regarded as the definition of the measurement rate $\Gamma_m$.

If a measurement occurs in a time step, we randomize over which and how many sites  undergo measurements. This is in concurrence with the assumption of performing independent and Markovian measurements at each site. The number of sites that get measured are chosen assuming a Bernoulli distribution. Thus, given $N/2$ sites (where $N$ is the number of Majorana fermions), the probability $p(n)$ for $n$ sites to be projectively measured is
\begin{eqnarray}
    p(n) = \binom{N/2}{n} p_m^n (1-p_m)^{N/2-n},
    \label{eqn:p_m}
\end{eqnarray}
which can be considered the definition of the measurement probability $p_m$. We again take the standard Monte Carlo approach, with bin sizes being determined by Eqn.~\eqref{eqn:p_m}. The sites to be measured are also chosen randomly with each site having an equal probability of being measured. 

The probabilities corresponding to all possible measurements are calculated. Each site can be projected to an eigenstate of $\sigma^z_{i}$ via $\{P_{i}^{(k_i)}\}_{k_i\in\{0,1\}}$, where $P_{i}^{(k_i)\,2}=P_{i}^{(k_i)}$ and $\sum_k P_{i}^{(k_i)}=\mathbb{I}_i$. Here, $i$ labels the site and $k_i\in\{0,1\}$ labels the eigenstates of $\sigma^z_i$. Let $n$ measurements occur with outcomes listed in a measurement record (bit string) $\Vec{r}\equiv\langle r_1, r_2,\dots,r_n\rangle\in\{0,1\}^n$, where $r_\ell$ is the measurement outcome at the $\ell$th measured site. We define the total projector corresponding to the measurement record $\Vec{r}$ as
\begin{equation}
    P^{(\Vec{r})}\equiv\prod_{i=1}^n P_{i}^{(r_i)}.
\end{equation}

Consider the state $\Psi(t_1;\mathcal{J})$ which has evolved under the Hamiltonian $H_{\mathcal{J}}$ for time $t_1$ per Eqn.~\eqref{eq:rho_HJt} but has \textit{not} previously been measured. The state after the first set of $n$ local measurements with outcomes $\Vec{r}_1$ is then, 
\begin{eqnarray}
    \Psi(t_1;\Vec{r}_1, \mathcal{J}) = \frac{P^{(\Vec{r}_1)}\Psi(t_1;\mathcal{J})P^{(\Vec{r}_1)}}{\Tr\Big(P^{(\Vec{r}_1)}\Psi(t_1;\mathcal{J})\Big)}.
    \label{eqn:meas}
\end{eqnarray}
The state $\Psi(t_1;\Vec{r}_1, \mathcal{J})$ is a quantum trajectory associated with the measurement result $\Vec{r}_1$ at time $t_1$ that is realized with probability $p(\Vec{r}_1|\mathcal{J})\equiv\Tr\big(P^{(\Vec{r}_1)}\Psi(t_1;\mathcal{J})\big)$; note that this necessarily depends on the Hamiltonian realization $H_{\mathcal{J}}$ as well. Following the first round of measurements, the dynamics for times $t>t_1$ depend on the previous measurement record and are determined by interlacing deterministic Hamiltonian evolution with further measurements. If $K$ rounds of measurements occur at times $t_1,t_2,\dots,t_K$ (with average spacing $t_{i+1}-t_i\approx1/\Gamma_m$) with outcomes $\Vec{r}_1,\Vec{r}_2,\dots,\Vec{r}_K$, then we describe the entire measurement history via $\Vec{R}\equiv\bigoplus_{i=1}^K\Vec{r}_i$. We formally write the associated probability for the measurement history $\Vec{R}$ as $p(\Vec{R}|\mathcal{J})$.

All things considered, the monitored dynamics of the SYK chain has three primary parameters which dictate global properties of the system: 
\begin{enumerate}
    \item The entanglement growth rate $\Gamma_{\rm{egr}}$ [Eqn. \eqref{eqn:gamma_egr}] determines how fast correlations spread and how fast the state returns to its unperturbed dynamics; $\Gamma_{\rm{egr}}$ depends on the interaction strength $J$ (Fig.~\ref{fig:EGR}).
    \item The measurement rate $\Gamma_m$ [Eqn. \eqref{eqn:gamma_m}] gives the rate at which measurements (perturbations) occur. 
    \item The measurement probability $p_m$ [Eqn. \eqref{eqn:p_m}] conveys the strength of the measurements, in the sense that higher values of $p_m$ leads to a larger chunk of the SYK chain getting projected onto a product state.
\end{enumerate}
To gain more insight about the dynamics, note that we can interpret $\Gamma_m$ as the coupling rate between the SYK chain and a `bath' of measurement devices. Given the likelihood of a measurement is $p_m$, this results in an effective (average) decoherence rate $\Gamma^{(\rm eff)}_{m}\equiv p_m\Gamma_m$ due to coupling to the bath. This interpretation is similar to the Lindblad approach taken in Ref.~\cite{pichler2022entanglemenTraj} for analyzing the open-system dynamics of a random Brownian circuit for a chain of spin-$1/2$ particles. The authors of~\cite{pichler2022entanglemenTraj} introduced a homodyne tuning parameter $\varphi$ which governs the non-unitary part of the evolution and thus the unraveling into particular quantum trajectories; here, the measurement probability $p_m$ has a similar function (in particular, $p_m\sim\cos^2\varphi$). 

To evaluate typical behavior, we generally focus on ensemble averages of quantities, such as entropy and purity, where the average is over the random couplings $\mathcal{J}$ and measurement histories $\Vec{R}$. Consider a particular quantum trajectory $\Psi(t; \vec{R},\mathcal{J})$ at time $t$ and a function of the trajectory $f(t;\Vec{R},\mathcal{J})$, which may be non-linear in $\Psi$. We formally define the ensemble average of $f$ as 
\begin{equation}\label{eq:ensemble_avg}
    \dblval{f(t)}\equiv\sum_{\Vec{R}} \int\dd{\mathcal{J}}p\big(\vec{R}|\mathcal{J}\big)p(\mathcal{J})f\big(t;\Vec{R},\mathcal{J}\big),
\end{equation}
where the sum is over all possible measurement histories $\Vec{R}$ [with history $\Vec{R}$ occurring with probability $p(\Vec{R}|\mathcal{J})$] and the integral is over the random couplings $\mathcal{J}$. We adopt double-bracket notation throughout to convey that the average is with respect to two random variables. Note that the average depends on the coupling $J$, the measurement probability $p_m$, and the measurement rate $\Gamma_m$ as well as the initial state. 


\begin{figure}
    \centering
    \includegraphics[width=.9\linewidth]{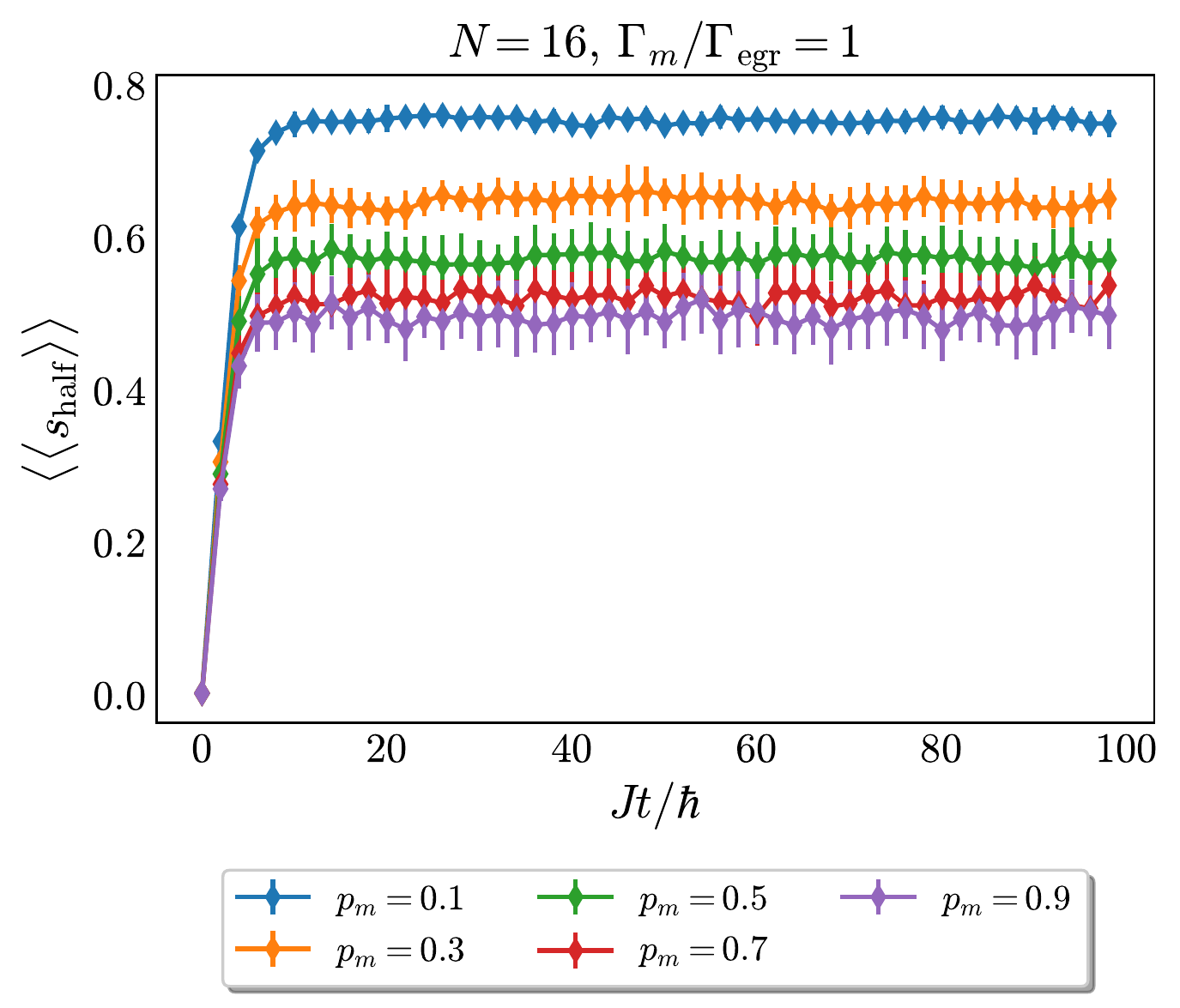}
    \includegraphics[width=.9\linewidth]{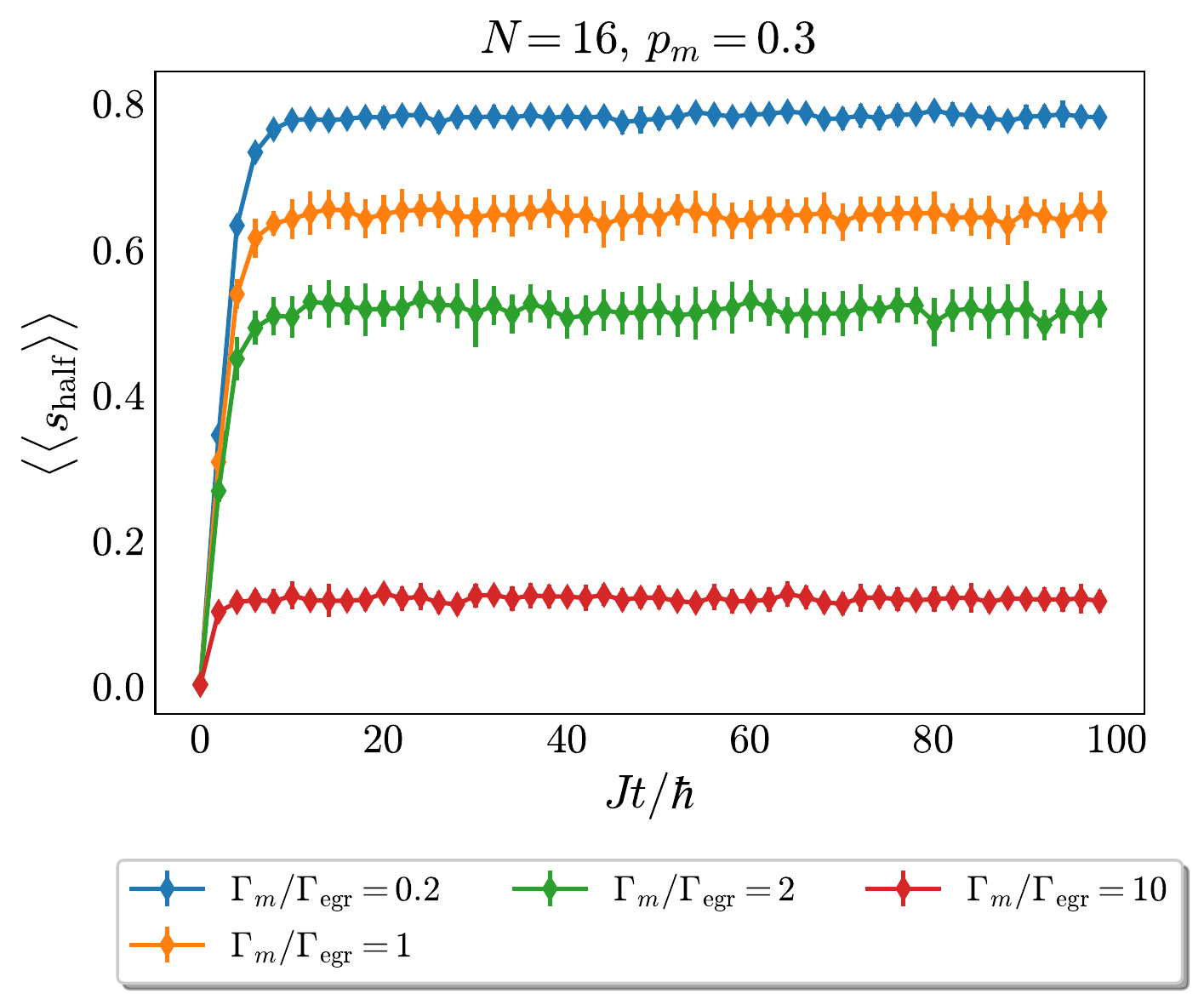}
    \caption{Entanglement dynamics of the SYK chain. (Top) Half-chain entanglement entropy for different values of measurement probabilities $p_m$ with a fixed measurement rate $\Gamma_m/\Gamma_{\rm egr}=1$.
    (Bottom) Half-chain entanglement entropy for different values of $\Gamma_m/\Gamma_{\rm egr}$ and fixed $p_m=.3$. Dynamics have been averaged over 20 batches with 50 runs each. Error bars show the standard deviation from the batches.}
    \label{fig:S_half_dynamics}
\end{figure}

\section{Results}
We present our numerical results of MIPTs for the SYK chain with $N=16$ Majorana fermions. Recall that a MIPT refers to global changes of a system's many-body quantum state---such as many-body entanglement or global purity of the system---induced by continuously monitoring the system. An entanglement MIPT refers to a transition from a state with extensive entanglement entropy (in the volume-law phase) to a state with sub-extensive entanglement entropy (in the area-law phase). Whereas a purification MIPT---quantified by the purity of the many-body state---refers to a transition from a highly mixed state (in the mixed phase) to a pure state (in the pure phase). As our numerical results here indicate, entanglement MIPTs and purification MIPTs are two distinct phenomena. In both cases however, a competition between internal scrambling dynamics and decoupling dynamics governs which phase the system relaxes to.

Though there have been many measures to diagnose a MIPT---such as the entropy, Renyi entropies, a reference qubit etc.~\cite{li2018zeno,skinner2019MIPT,fisher2019driven,szy2019weakMeas,jian2020criticality,gullans2020criticality,bao2020theoryof,choi2020qec_prl,gullans2020purification,gullans2020scalable,fan2021self,li2021SM_qec}---we find the half-chain entanglement entropy and the purity of the global state sufficient to diagnose entanglement MIPTs and purification MIPTs, respectively. We thus focus on these two quantities throughout.

\subsection{Entanglement phase transition}

We compute the half-chain entropy for a single Hamiltonian realization $H_\mathcal{J}$ and measurement history $\Vec{R}$ at time $t$ starting from an initial pure state $\Psi(0)$ with all spins pointing up (${\Psi(0)=\dyad{1}^{\otimes N/2}}$). The entropy corresponds directly to the amount of entanglement within the system for a given $\mathcal{J}$ and $\Vec{R}$. We then average over many realizations and measurement histories to quantify the average entanglement entropy. Explicitly, given a quantum trajectory $\Psi(t;\Vec{R},\mathcal{J})$ at time $t$ described by measurement history $\Vec{R}$ and Hamiltonian $H_{\mathcal{J}}$, the (average) half-chain entanglement entropy at time $t$ is $\dblval{s_{\rm half}(t)}$ per Eq.~\eqref{eq:ensemble_avg}.

A competition between entanglement growth and decoupling (due to measurements) determine the entanglement dynamics of the SYK chain. If the measurements are too frequent (large $\Gamma_m/\Gamma_{\rm egr}$) and too strong (large $p_m$), then the system will have sub-extensive entanglement in the steady state (area-law phase); whereas if the converse is true, then the state will have an extensive amount of entanglement (volume-law phase). In Fig. \ref{fig:S_half_dynamics}, we plot the entanglement dynamics for different values of $p_m$ and $\Gamma_m$. Clearly, the entanglement entropy saturates to a lower values as the measurement probability $p_m$ increases; we observe similar effects when increasing the measurement rate $\Gamma_m$. For the latter, subtle hints of an entanglement MIPT can be observed when passing between $\Gamma_m \lesssim \Gamma_{\rm{egr}}$ and $\Gamma_m\gtrsim \Gamma_{\rm{egr}}$.

\begin{figure}
    \centering
    \includegraphics[width=.9\linewidth]{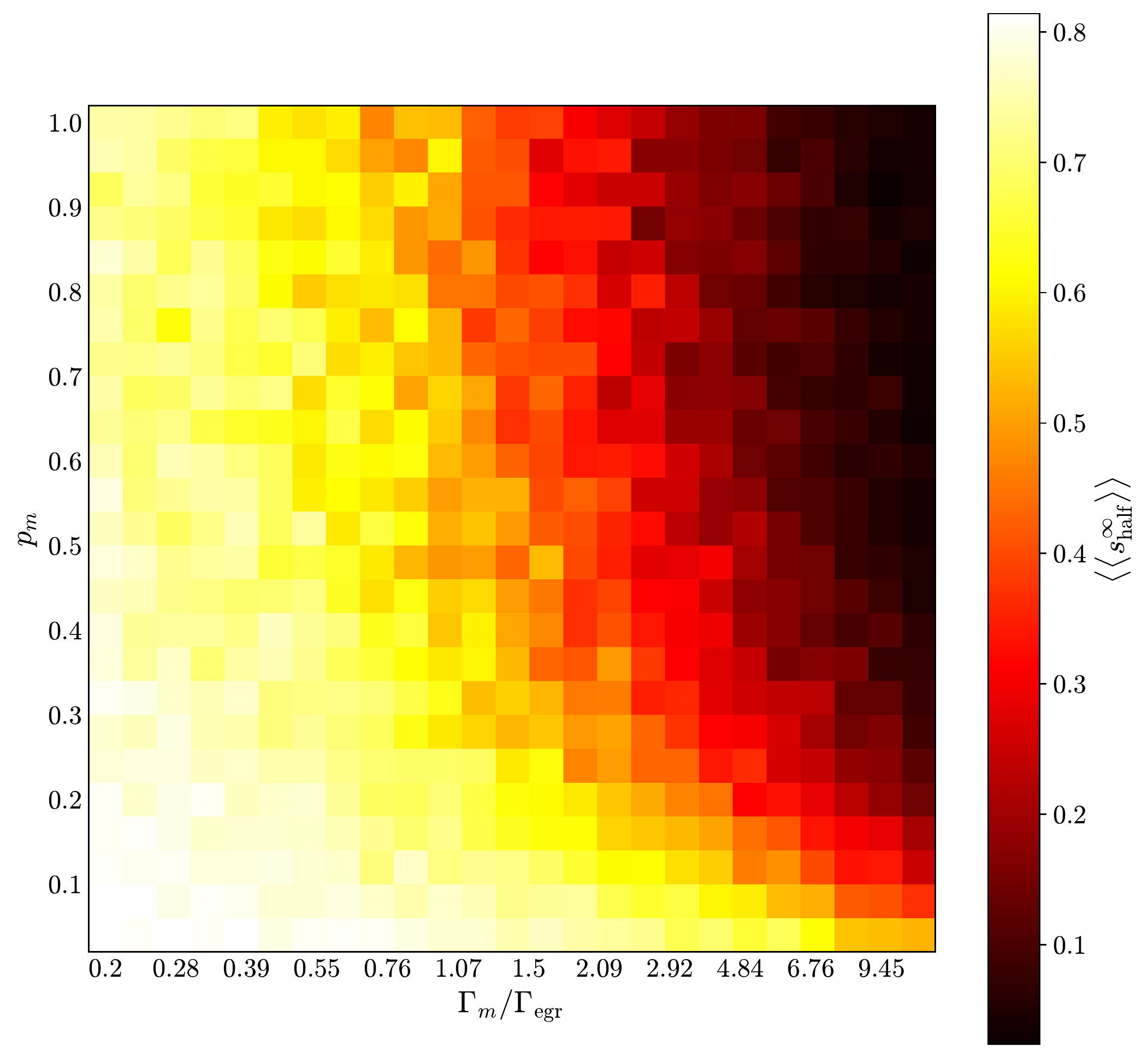}
    \caption{Entanglement phase diagram. Steady-state values of the half-chain entanglement entropy as a function of measurement probability $p_m$ and measurement rate $\Gamma_m/\Gamma_{\rm egr}$. Here, $J=1$ ($\Gamma_{\rm egr}=.2)$, $N=16$, and a steady-state time $t_\infty=200$ is chosen. All dynamics have been averaged over 50 runs.}
    \label{fig:S_half_saturation}
\end{figure}

To more clearly highlight the emergence of an entanglement MIPT, we look at the steady-state entanglement entropy $\dblval{s_{\rm half}(\infty)}$. Practically, due to finite-time simulations and finite-size effects, we cannot go to the $t\rightarrow\infty$ limit. Instead, we pick a large enough time $t_{\infty}\gg(\Gamma_m^{-1},\Gamma_{\rm egr}^{-1})$ to observe relaxation but not too large so that finite-size effects become significant. We then compute $\dblval{s_{\rm half}(\infty)}$ for various values $(\Gamma_m/\Gamma_{\rm egr},p_m)$ with fixed EGR $\Gamma_{\rm egr}\approx.2$ ($J=1$; see Fig.~\ref{fig:EGR}). We plot the results in a 2D phase-diagram in Fig.~\ref{fig:S_half_saturation} and witness an entanglement MIPT---from extensive entanglement (yellow-white) to sub-extensive entanglement (red-black)---as $\Gamma_m$ and $p_m$ increase. The fuzzy region in between is likely due to finite-size effects, as we only simulate $N/2=8$ spins. Nevertheless, from the diagram, we see that, for each value $0<p_m\leq1$, there exists a critical measurement rate $\Gamma_m^c$ where a MIPT occurs. Interestingly, even at $p_m=1$, there is extensive entanglement in the steady-state if $\Gamma_m$ is low enough. Similar phenomenon was found in recent studies on random Brownian circuits~\cite{pichler2022entanglemenTraj} and (randomly) coupled SYK chains in the large $N$ limit~\cite{swingle2022longrangeMIPT}. This is intriguing because, for $p_m=1$, the many-body state frequently (on a time scale $T\sim1/\Gamma_m$) gets projected into a product state of the form $\bigotimes_{i=1}^{N/2}\dyad{k_i}$, where $k_i\in\{0,1\}$ is the measurement outcome at the $i$th site, however the many-body correlations quickly revives so that the system spends the majority of the time in a highly entangled state. We note that such a revival (and thus the appearance of a nontrivial $\Gamma_m^c$ at $p_m=1$) is distinct to entanglement dynamics and does not occur in purification dynamics, as we discuss below.

\subsection{Purification phase transition}
We now analyze the purification dynamics of an initially, maximally mixed (infinite temperature) state $\rho_0=\mathbb{I}/2^{N/2}$; i.e., after a time $t$ of evolution, we compute the ensemble averaged purity $\dblval{\Tr{\rho^2(t)}}$ per Eqn.~\eqref{eq:ensemble_avg}, where $\Tr{\rho^2(0)}\leq\dblval{\Tr{\rho^2(t)}}\leq1$ and ${\Tr{\rho^2(0)}}=1/2^{N/2}$.  

\begin{figure}
    \centering
    \includegraphics[width=.9\linewidth]{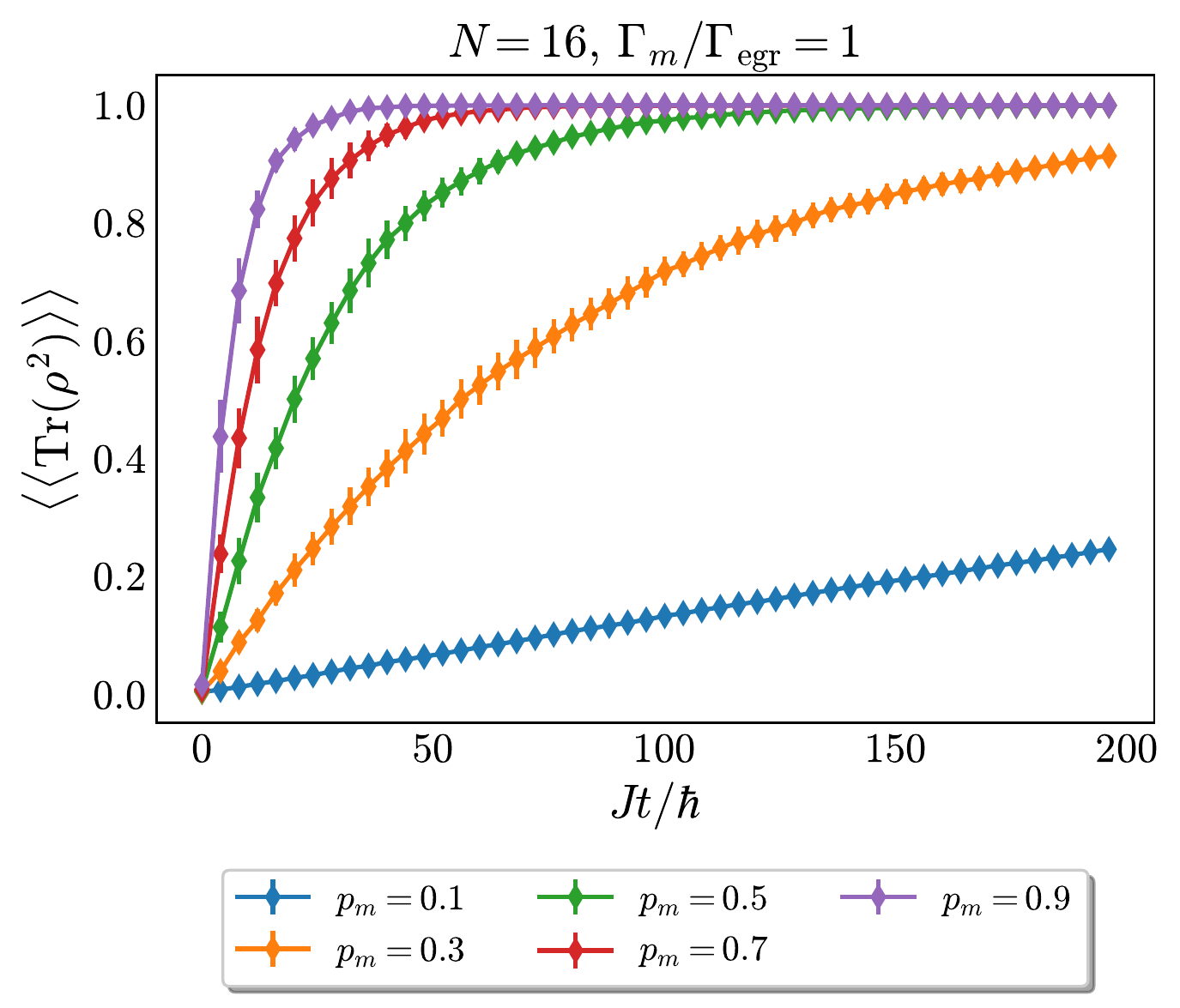}
    \includegraphics[width=.9\linewidth]{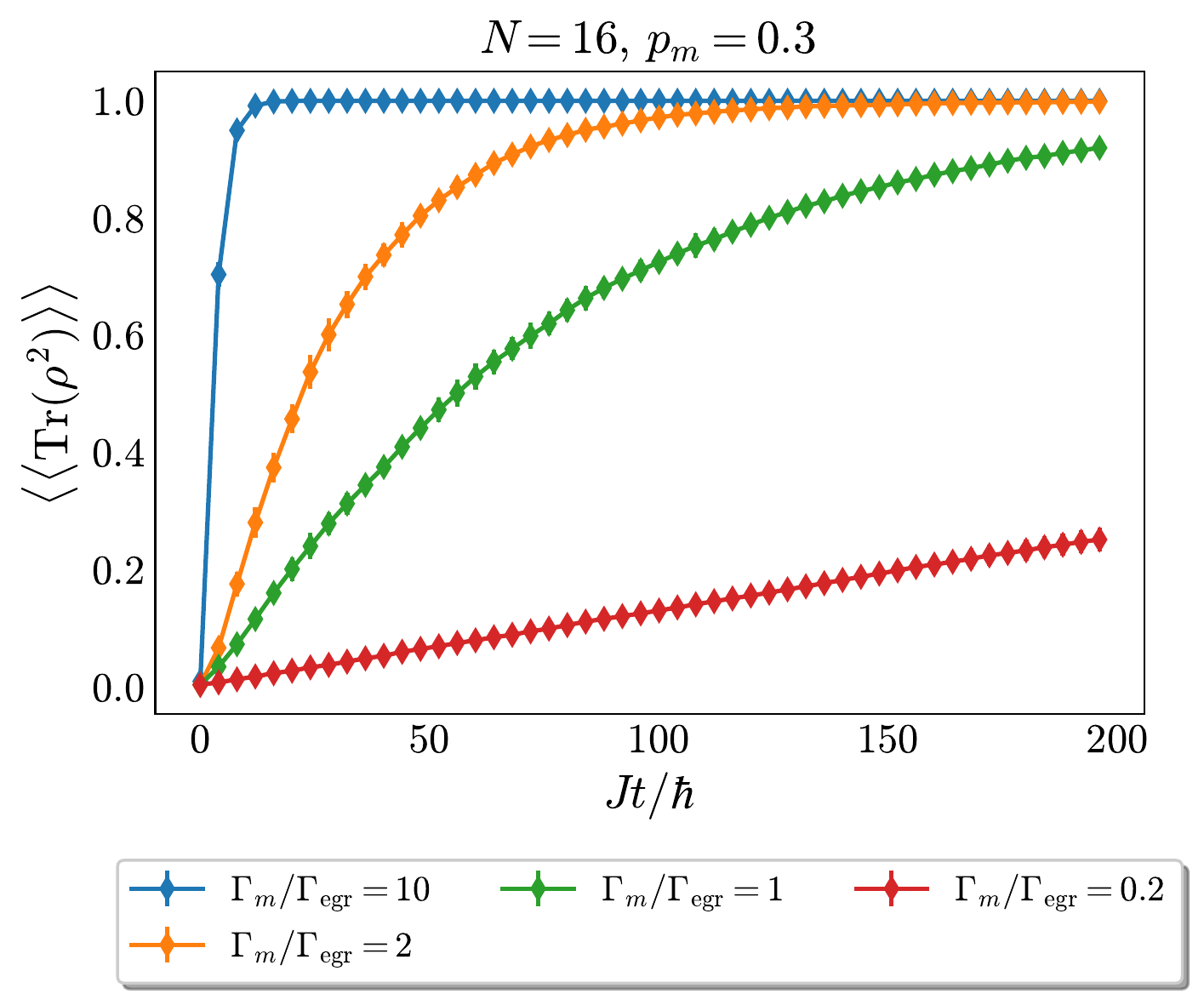}
    \caption{Purification dynamics of the SYK chain. (Top) Purity for different values of measurement probabilities $p_m$ with a fixed measurement rate $\Gamma_m/\Gamma_{\rm egr}=1$. (Bottom) Purity for different values of $\Gamma_m/\Gamma_{\rm egr}$ and fixed $p_m=.3$. Dynamics have been averaged over 20 batches with 50 runs in each. Error bars show the standard deviation over the 20 batches.}
    \label{fig:Purity_dynamics}
\end{figure}

Measurements (strictly) increase purity and localize the remaining impurity to complementary regions which have not yet been measured. Internal dynamics, however, scrambles the impurity into many-body correlations of the system, reducing the purifying effects of later measurements. A competition arises between scrambling and purification leading to different purification phases---a mixed phase and a pure phase---depending on whether scrambling or measurement dominates. In Fig.~\ref{fig:Purity_dynamics}, we plot the purification dynamics for different values of $p_m$ and $\Gamma_m$. For high values of the measurement strength $p_m\approx1$, the system necessarily purifies on time-scales $\sim1/\Gamma_m$, however for intermediate values $0<p_m<1$, the system can either purify or remain mixed for times $T\gg1/\Gamma_m$ depending on whether $\Gamma_m/\Gamma_{\rm egr}\gg1$ or $\Gamma_m/\Gamma_{\rm egr}\ll1$, as seen qualitatively in Fig.~\ref{fig:Purity_dynamics}. This is indicative of a purification MIPT.

\begin{figure}
    \centering
    \includegraphics[width=.9\linewidth]{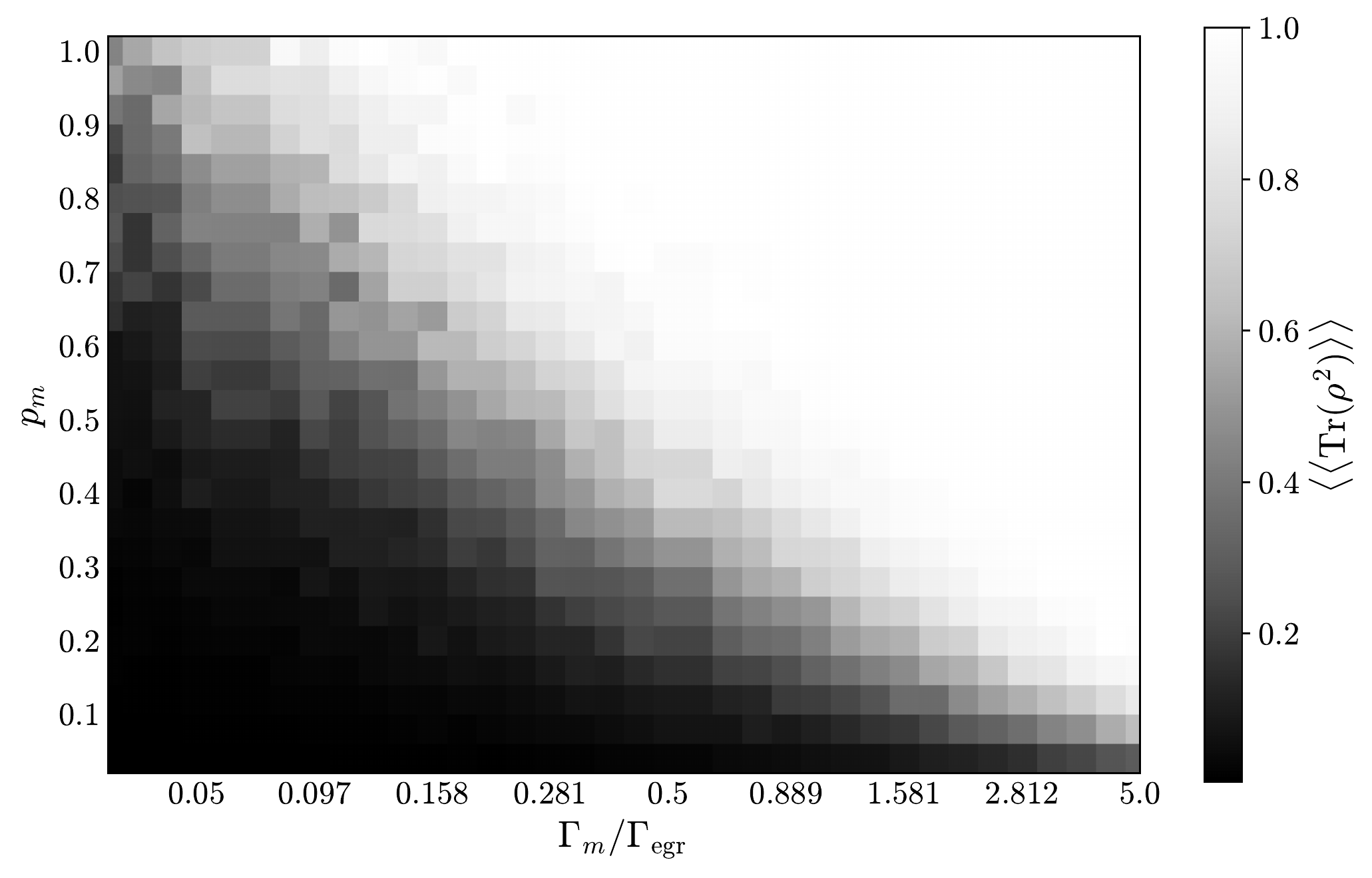}
    \caption{Purification phase diagram. Steady-state values of the purity as a function of measurement probability $p_m$ and measurement rate $\Gamma_m/\Gamma_{\rm egr}$. Here, $J=1$ ($\Gamma_{\rm egr}=.2)$, $N=16$, and a steady-state time $t_\infty=1000$ is chosen. All dynamics have been averaged over 50 runs.}
    \label{fig:purity_saturation}
\end{figure}

To more clearly highlight the emergence of a purification MIPT, we compute the purity in the steady-state $\dblval{\Tr{\rho^2(\infty)}}$ and plot the results in 2D parameter space $(\Gamma_m/\Gamma_{\rm egr},p_m)$ in Fig.~\ref{fig:purity_saturation} for fixed $\Gamma_{\rm egr}$. There is a clear demarcation (though with a fuzzy boundary due to finite-size effects) between the mixed phase (black region) and pure phase (white region). Thus, for a fixed measurement strength $p_m$, we can increase the measurement rate $\Gamma_m$ to go from the mixed phase to the pure phase (and likewise, we can tune $p_m$ starting from a fixed $\Gamma_m$). Unlike an entanglement MIPT, for $p_m=1$, any nonzero measurement rate will necessarily push the system to the pure phase within a time $\sim1/\Gamma_m$. This is actually general for any $p_m$ and $\Gamma_m$ since $\Tr{\rho^2(0)}\leq\dblval{\Tr{\rho^2(t)}}\leq1$ for any $p_m$, $\Gamma_m$, and time $t>0$. The inequality follows from the simple fact that unitary evolution and projective measurements cannot decrease purity. Such an inequality cannot be written for entangling dynamics since unitary evolution generally leads to an in increase in the entanglement entropy (for any bipartite cut across the system).

\subsubsection{Purification time-scales}

\begin{figure}
    \centering
    \includegraphics[width=.9\linewidth]{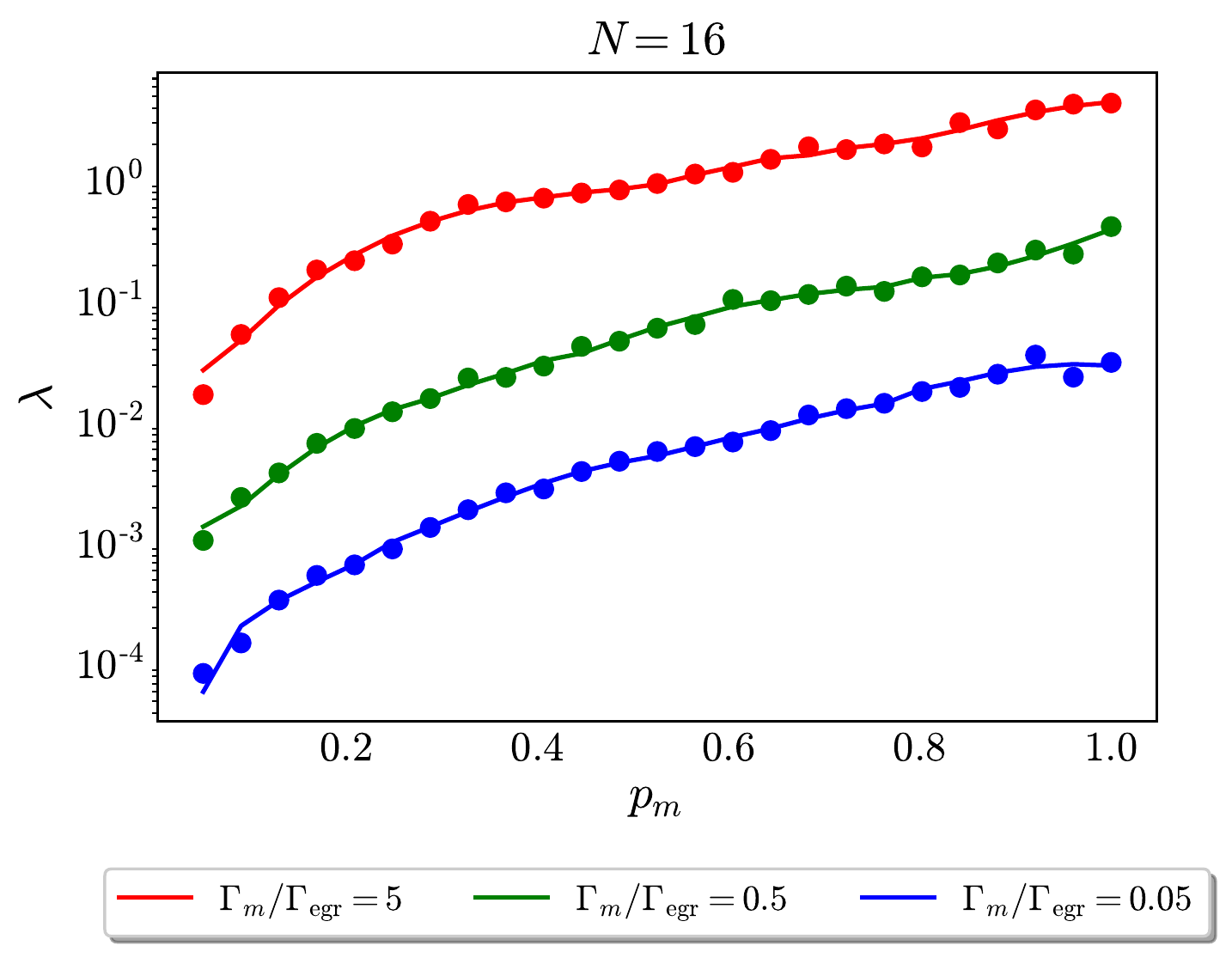}
    \caption{Purification rate $\lambda$ assuming the ansatz $\dblval{\Tr{\rho^2(t)}}\doteq\tanh(\lambda t + \alpha)$, where $\tanh\alpha = \Tr{\rho^2(0)}=1/2^{N/2}$. The purification rate is approximately linear in the measurement rate $\Gamma_m$. Trends show a strong non-linear dependence on $p_m$.}
    \label{fig:Purity_lambda}
\end{figure}

We attempt to find an intrinsic, purification time-scale directly from the data. Figure \ref{fig:Purity_dynamics} suggests that the dynamics of purity closely resembles a $\tanh$-like profile. We thus take the following ansatz
\begin{eqnarray}
    \dblval{\Tr{\rho^2(t)}} \doteq \tanh(\lambda t + \alpha),
\end{eqnarray}
with $\tanh\alpha=\dblval{\Tr{\rho^2(0)}} = 1/2^{N/2}$. Here, the purification time-scale is quantified by the fitting  parameter $\lambda$, which depends on $p_m$ and $\Gamma_m/\Gamma_{\rm egr}$. We find a good fit with $R^2$ values ranging between 0.996 and 0.999 for different choices of $p_m$ when $\Gamma_m/\Gamma_{\rm egr} = 5$, between 0.994 and 0.999 when $\Gamma_m/\Gamma_{\rm egr} = 0.5$ and between 0.983 and 0.997 when $\Gamma_m/\Gamma_{\rm egr} = 0.05$. 

From the fit, we extract the purification rate $\lambda$ for several values of $p_m$ and $\Gamma_m$ and plot the results in Figure \ref{fig:Purity_lambda}. We note that $\lambda$ is close to linear in $\Gamma_m$, as intuitively expected, and that $\lambda(p_m=1)\sim\Gamma_m$ up to some $\order{1}$ constant. Hence, $\lambda\sim\Gamma_m f(p_m)$ for some function $f(p_m)$ that may also depend on the critical strength $p_m^c$ and critical rate $\Gamma_m^c$. From the data in Fig.~\ref{fig:Purity_lambda}, we see that $f(p_m)$ shows strong non-linear behavior in $p_m$, with $f(p_m)$ changing over two orders of magnitude as the measurement strength $p_m$ is tuned from 0 to 1.


\section{Discussion}

\begin{figure}
    \centering
    \includegraphics[width=0.9\linewidth]{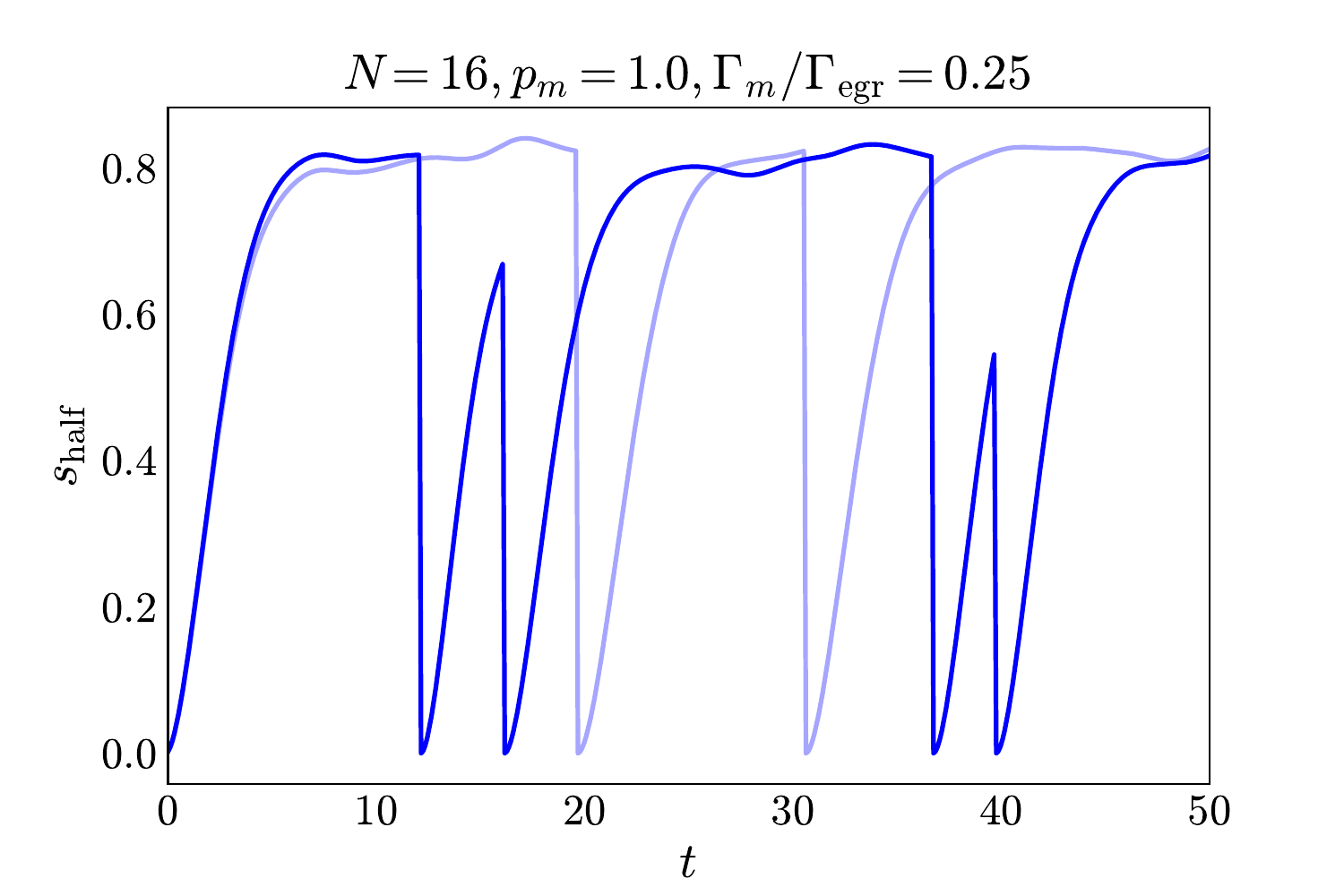}
    \includegraphics[width=0.9\linewidth]{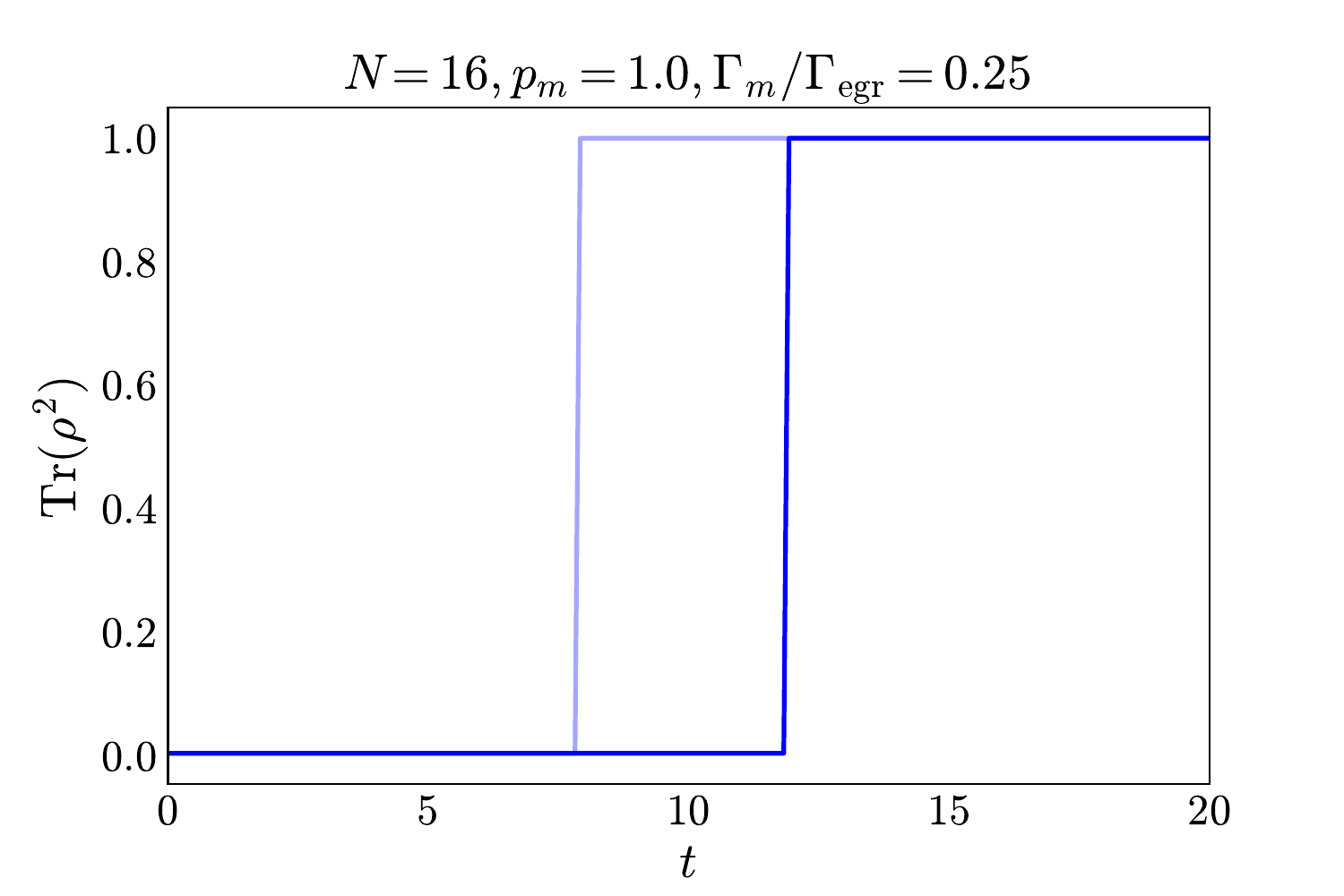}
    \caption{Quantum jumps in monitored dynamics. (Top) Revival of entanglement entropy and (Bottom) purification transition for two distinct quantum trajectories (dark blue, light blue) in the regime of completely projective ($p_m=1$) but relatively infrequent ($\Gamma_m/\Gamma_{\rm egr}=.25$) measurements.}
    \label{fig:s_revival}
\end{figure}

At this juncture, we point out an interesting viewpoint on purification phases as seen through the lens of quantum error correction~\cite{choi2020qec_prl,gullans2020purification,gullans2020scalable,ippoliti2021postselection,fan2021self,gullans2021low_depth,li2021SM_qec,hastings2021qumems}. We can think of the SYK chain as a quantum memory which is entangled (e.g., shares a large number of Bell pairs) with a quantum computer. The quantum error correction properties of the memory refer to how well the memory retains entanglement with the quantum computer in the face of external, deleterious perturbations or errors from the environment (e.g., measurement devices). The internal unitary dynamics of the memory acts as a quantum error correcting code that hides information (e.g., entanglement with the quantum computer) from the environment by scrambling the information non-locally into the memory's many degrees of freedom. In order to access this information, the environment must couple to an extensive number of memory qubits. Purification of the system is then interpreted as wiping the memory of its initial entanglement with the quantum computer; see Appendix~\ref{sec:decoupling} for further discussion on this perspective.

Entanglement dynamics does not admit an equivalent QEC description because the half-chain entanglement entropy captures both the initial information encoded in the system as well as the many-body correlations that build up over time, and these two quantities are not mutually inclusive. On the other hand, it is often said in the literature that entanglement and purification MIPTs are two sides of the same coin, but this is an over-simplification and is misleading. From previous discussions and by comparing the phase diagrams of Figs.~\ref{fig:S_half_saturation} and~\ref{fig:purity_saturation}, we see that entanglement MIPTs and purification MIPTs are in fact two distinct phenomena. The reason for this distinction is quite simple. For an entanglement MIPT, there is a revival of entanglement even after the system has been completely projected onto a product state ($p_m=1$); see Fig.~\ref{fig:s_revival}. We can thus have extensive entanglement entropy in the steady state---even though the system has lost all information about initial conditions---if the entanglement growth is faster than the rate of measurements ($\Gamma_m/\Gamma_{\rm egr}<1$) because the system spends most of its time in highly entangled states and is only projected here-and-there into product states.\footnote{The steady-state is fluctuating (and thus not quite steady) due to random quantum jumps induced by measurements, however one can coarse-grain over a time-scale $\sim1/\Gamma_m$ to witness steady behavior. This is effectively done by our averaging procedure since the measurement process is a Poisson process here.} Quantitatively, this leads to a non-trivial critical measurement rate $\Gamma_m^c\sim\Gamma_{\rm egr}$ at $p_m=1$; for $\Gamma_m\gtrsim\Gamma_m^c$, the system does not have time to recover to a highly entangled state before another completely projective measurement occurs, and the system spends the majority of the time in a product state. Contrariwise, for a purification MIPT, there is no such revival of the mixed phase once the system transitions to the pure phase (Fig.~\ref{fig:s_revival}) because the purification MIPT signals a true loss of initial conditions; i.e., once the state is pure, it remains pure. Indeed, this follows directly from the inequality $\Tr{\rho^2(0)}\leq\dblval{\Tr{\rho^2(t)}}\leq1$ which holds for monitoring dynamics. Hence the critical measurement rate for a purification MIPT at $p_m=1$ is trivial. Interpreting the system as a quantum memory, projections effectively wipe the memory within a time $\sim1/\Gamma_m$, and there is no ``rewriting'' into the memory thereafter.

\acknowledgements
SH and AJB have benefited from discussions with Vishal Katariya in the initial stages of this study. SH acknowledges financial support from the Army Research Office Multidisciplinary University Research Initiative (ARO MURI) through the grant number W911NF2120214. AJB acknowledges financial support from the Defense Advanced Research Projects Agency (DARPA) under the Young Faculty Award (YFA) Grant No. N660012014029.

\appendix

\section{Purification from decoupling}\label{sec:decoupling}

\begin{figure}[b]
    \centering
    \includegraphics[width=\linewidth]{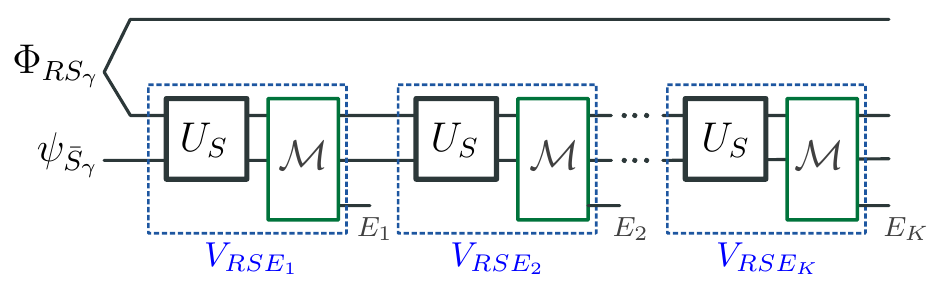}
    \caption{Quantum circuit view of purification dynamics. Part of the system $S$ is initially entangled with a reference $R$ such that the initial state of the system is in a mixed phase. Internal, unitary evolution $U_S$ entangles the system's constituents---spreading the initial impurity non-locally into many-body degrees of freedom---while measurements (with isometric extension $\mathcal{M}$) decouples the systems qubits; $U_S$ and $\mathcal{M}$ together give full isometric extension $V_{RSE_i}=\mathbb{I}_R\otimes V_{SE_i}$ for the $i$th measurement round, where $E_i$ labels the set of measurement devices.}
    \label{fig:isometric_meas}
\end{figure}

We can qualitatively and quantitatively examine the QEC properties of a scrambling system (e.g., SYK chain) undergoing continuous monitoring~\cite{choi2020qec_prl,gullans2020purification,gullans2020scalable,ippoliti2021postselection,fan2021self,gullans2021low_depth,li2021SM_qec,hastings2021qumems} by applying the so-called decoupling principle~\cite{schumacher2002approx,hayden2008decoupling,dupuis2010thesis}, which can be explained by the following example. Consider a system $S$ initially in a mixed state $\tau_S$ which admits a purification $\Psi_{RS}$ such that $\tau_S=\Tr_R(\Psi_{RS})$, where $R$ is a reference that is entangled with $S$. For instance, consider a system of $N$ qubits (such that $|S|=2^N$) with some fraction $\gamma N$ qubits in a maximally mixed state and the remaining $(1-\gamma)N$ qubits in some pure state---i.e., $\tau_S=\pi_{S_\gamma}\otimes\psi_{\bar{S}_\gamma}$ where $\pi_{S_\gamma}=\mathbb{I}/|S_\gamma|$ and $|S_\gamma|=2^{\gamma N}$ ($|\bar{S}_\gamma|=2^{(1-\gamma)N}$). Then one purification is $\Psi_{RS}=\left(\bigotimes_{i=1}^{\gamma N}\Phi_{R_iS_{\gamma_i}}\right)\otimes \psi_{\bar{S}_\gamma}$, where $\Phi_{R_iS_{\gamma_i}}$ is a Bell pair consisting of the $i$th qubit in $S_\gamma$ and the $i$th qubit in $R$. Now consider an isometric channel $\mathcal{V}:RS\rightarrow RSE$ that couples the system $S$ to an environment $E$ such that, 
\begin{equation}
    \Psi_{RSE}\equiv\mathcal{V}(\Psi_{RS})= V\Psi_{RS}V^\dagger,
\end{equation}
where $V^\dagger V=\mathbb{I}_{RS}$ and thus $\Psi_{RSE}$ is pure. We assume $V=\mathbb{I}_R\otimes V_{SE}$---i.e., the reference acts as a bystander; see Fig.~\ref{fig:isometric_meas} for an illustration. An instance of this general setup is the purification dynamics consisting of unitary evolution and measurements starting from the initially mixed state $\tau_S$, where $V$ encodes the internal unitary evolution as well as (an isometric extension of) the external measurements by measurement devices $E$.\footnote{The inclusion of measurements into the isometry is possible by the principle of deferred measurements; i.e., we coherently couple/entangle the measurement devices to the system and projectively measure the measurement devices at the end.} 

We are now in the position to state the decoupling principle. Given the initial state $\Psi_{RS}$ sent through the channel $\mathcal{V}$, the system $S$ and reference $R$ maintain the entanglement within the state $\Psi_{RS}$ with fidelity $1-\epsilon$ if the reference $R$ and environment $E$ are approximately in a product state (decoupled); i.e.,  
\begin{equation}
    \norm{\rho_{RE}-\rho_R\otimes\rho_E}_1\leq\epsilon\sim\order{\exp(-cN)},
\end{equation}
where $\norm{\sigma}_1=\Tr\,(\sqrt{\sigma\sigma^\dagger})$ is the trace norm, $\rho_{RE}=\Tr_S(\Psi_{RSE})$, $\rho_{E}=\Tr_{RS}(\Psi_{RSE})$, $\rho_{R}=\Tr_{SE}(\Psi_{RSE})$ and $c$ is some constant independent of $N$. 

For instance, given $\gamma N$ initial Bell pairs between $S$ and $R$ and a random, scrambling unitary $U_S$ followed by measurements on (randomly chosen) $p_m N$ system qubits, ${c=1-p_m-\gamma}$~\cite{choi2020qec_prl}. In this case, decoupling fails as $\gamma\rightarrow1-p_m$. Observe that $\gamma$ quantifies the initial purity of the system via $\Tr(\tau_S^2)=1/2^{\gamma N}$. Let $\gamma=1/N$ such that there is initially only one bit of impurity in the system [$\Tr(\tau_S^2)=1/2$]. In the thermodynamic limit ($N\rightarrow\infty$) the system remains impure for all $p_m\leq 1$, whereas a purification phase transition occurs at $p^c_m=1$~\cite{choi2020qec_prl}. For unitaries $U_S$ that do not completely scramble within a time $1/\Gamma_m$, the critical point occurs at lower values $p^c_m<1$, which is the case for MIPTs in low-depth brickwork circuits (for which $p_m\approx.16$~\cite{li2018zeno,skinner2019MIPT,fisher2019driven,szy2019weakMeas,jian2020criticality,gullans2020criticality,bao2020theoryof}).

The analysis above is for a single round of measurements, however the purification phase is stable in the steady state (i.e., after a large number of successive, iid measurements) for reasonable time scales, as we show in Figs.~\ref{fig:Purity_dynamics} and~\ref{fig:purity_saturation} of the main text for an SYK chain. To see this from a decoupling perspective, consider $K$ rounds of iid measurements (interlaced with random, internal unitary evolution), with each successive measurement occurring at a rate $\Gamma_m$, and consider the associated isometric channel $\mathcal{V}^{(K)}: RS\rightarrow RSE^K$, where $E^K$ denotes the set of measurement devices for $K$ measurement rounds. In particular, the isometry is $V^{(K)}=\mathbb{I}_R\otimes(\bigotimes_{i=1}^K V_{SE_i})$, where $E_i$ refers to the measurement devices for the $i$th round. Define the output state of the isometric channel as $\Psi_{RSE}^{(K)}\equiv\mathcal{V}^{(K)}(\Psi_{RS})$ where $\Psi_{RS}$ is a pure input. Then, the system $S$ and reference $R$ maintain the entanglement within the state $\Psi_{RS}$ with fidelity $1-K\alpha\epsilon$, where $\alpha\sim\order{1}$ constant, if the reference $R$ and the environment $E^K$ are approximately decoupled; i.e.,
\begin{equation}\label{eq:stable_ineq}
    \norm{\rho_{RE}^{(K)}-\rho_R\otimes\rho_E^{\otimes K}}_1\leq K\alpha\epsilon,
\end{equation}
where $\rho_E^{\otimes K}=\bigotimes_{i=1}^K\rho_{E_i}$ and $\rho_{E_i}$ is the state of the $i$th environment (i.e., the $i$th set of measurement devices) after the $i$th measurement round. Why should we expect Eqn.~\eqref{eq:stable_ineq} to hold? The reason being that the total error is bounded by the sum of errors in each step. [This can be shown explicitly by successively applying the triangle inequality ${\norm{(a-b) +(b-c)}_1\leq\norm{a-b}_1+\norm{b-c}_1}$ to the left hand side of Eq.~\eqref{eq:stable_ineq}.] In turn, the error per step scales as $\epsilon$ up to some $\order{1}$ constant such that the average error is $\alpha\epsilon$.

From Eqn.~\eqref{eq:stable_ineq}, we have that, for $K\ll1/\epsilon$, the reference $R$ and environment $E$ remain decoupled. In the context of purification phases, the system remains in the mixed phase for all reasonable timescales $T\sim{\rm poly}(N)/\Gamma_m$, where $\Gamma_m$ dictates the frequency of measurements. This gives some credence to the long-time stability of the mixed phase shown in Fig.~\ref{fig:Purity_dynamics}.

%

\end{document}